# Overhead Control with Reliable Transmission of Popular Packets in Ad-Hoc Social Networks

Feng Xia, *Senior Member, IEEE,* Hannan Bin Liaqat, Jing Deng, *Senior Member, IEEE,* Jiafu Wan, Sajal K. Das, *Fellow, IEEE*

*Abstract*— Reliable social connectivity and transmission of data for popular nodes is vital in multihop Ad-hoc Social Networks (ASNETs). In this networking paradigm, transmission unreliability could be caused by multiple social applications running on a single node. This leads to contentions among nodes and connection paths. In addition, congestions can be the result of multiple senders transmitting data to a single receiver and every sender waiting for a positive acknowledgment to move on. Therefore, traditional Transmission Control Protocol (TCP) performs poorly in ASNETs, due to the fact that the available bandwidth is shared among nodes using round trip time and the acknowledgment is provided individually to every data packet. To solve these issues, we propose a technique, called Overhead Control with Reliable Transmission of Popular Packets in Ad-Hoc Social Networks (RTPS), which improves transmission reliability by assigning bandwidth to users based on their popularity levels: extra bandwidth is assigned to the nodes with higher popularity and their acknowledgments are sent with higher priority. In addition, RTPS further reduces contentions and packet losses by delaying acknowledgment packet transmissions. Our detailed investigations demonstrate the excellent performance of RTPS in terms of throughput latency and overhead with different hop-distances and different numbers of concurrent TCP flows.

## I. INTRODUCTION

COMMUNICATIONS between nodes in Ad-hoc Social Networks (ASNETs) depend on the nodes' social features such as social graph, human mobility pattern, similarity, centrality, community, and tie strength [1]. However, with increasing growth of social networks, reliable and congestion-free communications become difficult due to their complexity and dynamic nature of user movement [2]. The scarce bandwidth- as well as the hidden/exposed node problems in ASNETs lead to traffic congestions [3][4], thus causing reliability issues.

F. Xia and H. B. Liaqat are with School of Software, Dalian University of Technology, China.
J. Deng is with Department of Computer Science, University of North Carolina at Greensboro, USA
J. Wan is with School of Mechanical and Automotive Engineering, South China University of Technology, China
S. K. Das is with Department of Computer Science, Missouri University of Science and Technology, USA
Corresponding author: Feng Xia; (e-mail: f.xia@ieee.org)

Usually, Transmission Control Protocol (TCP) is considered as one of the best solutions to achieve reliability with congestion control [5]. However, TCP suffers from a number of known problems in multihop ad-hoc networks [6]. There exist many works trying to address these issues [7] [8]. Nevertheless, such schemes do not take into consideration the social properties. Furthermore, hidden/exposed nodes make the rate adjustment a huge challenge in ASNETs [9].

The reliability of sender data packet transmissions is affected by two main aspects: 1) when multiple senders are sending data on multihop ad-hoc networks and only a single receiver is receiving data, some of these packets are dropped close to the receiver, and 2) the acknowledgment packets might be lost when data and acknowledgment packets use the same path (the sender of the acknowledgment packets). Both of these types of scenarios would lead to the TCP module at the senders to react improperly.

An example of such problems in ASNETs is shown in Fig. 1 with three different Social Communities ($SC_A$, $SC_B$ and $SC_C$). Sender nodes 1, 2, and 3 reside in $SC_A$ and intend to send data to the receiver node 7 in $SC_C$. The transfer of data packets depends on the social property (popularity) of nodes. The major issues in our ASNET scenario are: Node 7 in $SC_C$ works as a single receiver that has lower bandwidth due to interfering nodes and lots of data packets received from multiple senders. Therefore, the receiver node cannot share the bandwidth reliably and fairly with all sender nodes. The existing solutions solve this problem by using the TCP scheme at the sender side, which is to divide bandwidth among users according to the Round Trip Time (RTT) mechanism [10]. However, this solution is not suitable for ASNETs where the assignment of bandwidth is based on social popularity of nodes. In addition, when lots of acknowledgments are sent back to the senders, another overhead issue occurs due to the collision between the data and acknowledgment packets after using the same path [11][12].

Fig. 1. ASNET with popular and non-popular nodes

To address the aforementioned problems, we propose an Overhead Control with Reliable Transmission of Popular Packets in Ad-Hoc Social Networks, called RTPS. Our scheme assigns more bandwidth to those more popular nodes. RTPS further reduces control overhead by delaying the acknowledgment packets that are contributing to the traffic congestions near the more popular nodes. To calculate the popularity level of the source node, RTPS uses the *degree centrality* concept that is based on the number of relationships with the source node. RTPS provides solution in conventional ad-hoc networks that exploits social property of each node for efficient utilization of resources [13]. In conventional ASNETs, where lot of users communicate and share resources with each other after using some social applications, RTPS also gives reliable communication to these nodes.

The main contributions of this paper are summarized as follows (Note that a very preliminary version of RTPS was proposed in [14]):

- RTPS supports higher reliability of data packets to more popular nodes by sharing bandwidth among sender nodes using popularity level, and secondly, it acknowledging earlier the data packets sent by popular sender nodes as compared to others.
- To reduce collision related losses in multi-hop ASNETs, RTPS adopts the delaying acknowledgment technique and sets delay acknowledgment window based on network conditions. The delay window releases acknowledgment packets before the expiration of Retransmission Time Out (RTO) at the sender side and sends acknowledgment after it receives out-of-order data packets at the receiver side.
- RTPS also provides least rate of bandwidth to less popular sender nodes. Furthermore, for efficient utilization of resources it shares bandwidth with the next popular node if the current popular node bears any high level of contention.
- RTPS totally depends on the transport layer with end-to-end semantic and it does not violate the layering concept. Furthermore, after adjustment of the rate at the receiver side it is helpful in designing TCP for socially-aware ad-hoc networks.
- We conduct extensive simulations to evaluate the performance of RTPS against other existing protocols and verify the efficiency of resources partitioned in nodes.

The rest of the paper is organized as follows. Related work is discussed in the next section. Section III describes a complete overview of RTPS and all the steps involved Section IV discusses the performance evaluation results while the conclusions are offered in Section V.

## II. RELATED WORK

A large number of users sharing a small amount of bandwidth lead to unreliability and congestions in ASNETs. To provide better communication with scarce resources, many approaches take advantage of user social features/properties [15]. In [16] the notion of *degree centrality* was used to address traffic congestion issues in ASNETs using social features, and a scheme termed BPD inspired from biological immune system is proposed in [17].

Social properties do not merely provide more advantages to utilize resources efficiently; they also provide reliability to some specific nodes, when lots of nodes communicate with each other. Therefore, we review below relevant literature in the context of reliability and congestion control.

### A. Opportunistic Network-based Reliability and Congestion Control Approaches

The movement of nodes in opportunistic networks is random with large delay and it depends on the human mobility pattern. As a result of large contact delay and unpredictable movement, it is difficult to route and disseminate data in this type of network. Different existing models for large delay networks with aspect of social property were discussed in [18]. Opportunistic network follows the Store Carry Forward (SCF) principle. It provides reliability through transferring data hop-by-hop and avoids congestion using buffer management.

To provide reliability in opportunistic networks, the authors in [19] presented four methods such as Custody Transfer (CT), Return Receipt (RR), CT notification, and bundle forwarding notifications. The consideration of CT and RR is suitable in scarce bandwidth and low-energy networks. This is because it uses less notification information as compared to CT notification and bundle forwarding notifications. In [20] was presented four methods employed in epidemic routing and offered reliability using hop-by-hop, active receipt, passive receipt and network bridge receipt concepts. The hop-by-hop method does not make significant contribution to reliability, while active receipt solves the reliability issue with high cost. Furthermore, the authors in [21] provide a reliable and efficient data delivery scheme that utilizes the concept of contact volume prediction in opportunistic network.

For solving the congestion issue in opportunistic networks, the single copy method has high priority because of less availability of resources. In this method, nodes delete messages from their own buffer after sending data to the next relay node. However, the single copy concept does not always solve the reliability issue in the network. The nodes involved in opportunistic networks use local information rather than global information which is hard to acquire. Therefore, the authors in [22] managed storage congestion using active opportunity cost. To optimize the overall revenue, the latter used revenue management and dynamic programming concept. Moreover, congestion issue can be solved in the multi copy case using replica management. In management policy, replica congestion is reduced in [23] by using a restricting method. A dynamic replication control method based on the network congestion condition is proposed in [24] while an extended version of CAFé, called CAFRep, with multiple copies of message is presented in [25]. CAFRep used social network properties, statistics of node buffer and ego networks to control the replication number of messages forwarded to the relay node.

### B. Reliability and Congestion Control Strategies in Ad-hoc Networks

The reliability of data packets in wireless ad-hoc networks is provided through TCP that gives reliability to senders after receiving an acknowledgment from the end node. Furthermore, based on the received acknowledgment, TCP avoids congestion related loss. A detailed survey involving host to host congestion control is presented in [26][27].

In order to ensure reliability and control acknowledgment in wireless multi-hop environments, we discuss the related work pertaining to TCP acknowledgment issues. Given the contradiction that sender nodes require acknowledgments from receivers to ensure reliability whereas network needs to delay acknowledgments for congestion control, to improve the performance of TCP, it was proposed in [28] that the receiver should delay acknowledgment until timeout occurs at the receiver side. To adjust the delay window dynamically, Al-Jubari *et al.* [29] presented TCP-ADW which uses inter-arrival time of packets at the receiver side to set the delay window. In [30], TCP-MDA uses TCP and MAC cross layer information to adjust the delay window. TCP-MDA adjusts delay window to a larger size when the collision probability is less and vice versa. It breaks the TCP end-to-end semantic due to cross layer communication and also requires to be evaluated in mobile networks. A cross layer solution called AP-DDA is proposed in [31] for Wireless Local Area Networks (WLANs). For a complete survey of improved end-to-end approaches reader may refer to [32].

In terms of reliability guarantee and congestion control, the above defined techniques are not suitable for ASNET scenarios where communications of nodes depend on social properties. In ASNETs, the reliability of data is provided through assigning extra bandwidth and sending acknowledgment earlier to the popular senders. This is because a popular node requires maximum sharing of data by utilizing extra bandwidth without occurrence of any losses. For assigning extra bandwidth and sending earlier acknowledgment to popular senders, RTPS considers the concept of *degree centrality*. In addition, our scheme avoids collision or reduces collision probability using delay acknowledgment techniques. RTPS sets the delay acknowledgment window based on the popularity level of senders. Furthermore, the size of acknowledgment window in RTPS is based on the estimation rate of each flow and network conditions.

TABLE I. DEFINITION OF NOTATIONS

| Notation | Definition |
|---|---|
| $e_k$ | Estimated rates of each flow |
| $l^{(c)}$ | Best desired link capacity |
| $m$ | $m$ is the total number of senders |
| $d_k^{(r)}$ | Desired rate of each sender node |
| $d_k^{(c)}$ | Degree centrality level of senders |
| $a_k$ | Advertised window at receiver side |
| $da_k$ | Delayed acknowledgment window |
| $w_k$ | Exponential weighted moving average rate |
| $c$ | Consumable link capacity of all nodes |
| $t_k$ | Arrival times of the current data packets |
| $l_k^{(r)}$ | Least rate for each sender node |
| $p_k^{(s)}$ | Current packet size of each source node |
| $e^{(r)}$ | Estimation ratio |
| $\ddot{e}$ | Increment and decrement factor |
| $t_k^{(s)}$ | Smoothed average arrival time |
| $t_k$ | Effective time-out interval |
| $\Delta_k$ | Estimated average rate difference |

### III. RTPS DESIGN FOR OVERHEAD AND RELIABILITY CONTROL

RTPS contains three major components: Link Capacity Computing Module (LCCM), Degree-centrality based Rate Calculation Module (DRCM) and Popularity-aware Flow and Acknowledgment Overhead Control Module (PFAOCM). We briefly describe the interactions among them with the help of Fig. 2 before detailing them in the following subsections.

Fig. 2 illustrates an ASNET with multiple Senders (S) communicating with a single Receiver (R) through wireless links. First, LCCM computes the estimated rates ($e_k$) for all senders ($k = 1$ *to* $m$, where $m$ is the total number of senders) and then calculates the best value for full utilization of link. After calculating the best desired link capacity ($l^{(c)}$), LCCM sends it to DRCM which in turn calculates the desired rates ($d^{(r)}$) based on $l^{(c)}$ and computes the *degree centrality ($d^{(c)}$)* for each of the sender nodes. Then the assigned $d^{(r)}$ values are sent to PFAOCM which sets the advertised window ($a$) and delayed acknowledgment window ($da$) based on the value of $d^{(r)}$ and exponential weighted moving average rate ($w$). Table. I explains the notations with their definitions we use over here.

### A. Link Capacity Computing Module

LCCM takes two steps to compute the optimum value of $l^{(c)}$. In the first step, it computes the estimated rate based on its observation of packet arrivals and arrival intervals. In the second step, it calculates the total number of data from all flows which is denoted as consumable link ($c$). Furthermore, for computation of each sender rate in our defined scenario, we assumed that all flows are long-lived such as file transfer.

First, we compute $e_k$, the estimated rate of a traffic flow [33]. The calculation of $e_k$ in RTPS follows the same idea as TCP-Jersey [33] but is processed at the receiver side:

$$e_k = \frac{p_k^{(s)} + \text{RTT} \cdot e_k^{(p)}}{\text{RTT} + (t_k - t_k^{(p)})} \qquad (1)$$

where $t_k^{(p)}$ and $t_k$ are the arrival times of the previous and current packets, respectively, $e_k^{(p)}$ is the previous estimated rate, $p_k^{(s)}$ is the current packet size, and RTT is the round-trip time. Here we use RTT at the receiver side; therefore, we use the concept of Timestamp option [34]. After calculating the average RTT value of each flow, RTPS utilizes the smoothed RTT of TCP.

In the second step, the main objective of LCCM is to fully utilize the available link capacity at receiver side by calculating the best desired link capacity of $l^{(c)}$. Furthermore, LCCM uses all the arrival rates $e_k$ to compute the utilization of link ($c$) through a simple summation of all $e_k$ from sender 1 to $m$.

In the initial stage of LCCM, the value of $l^{(c)}$ is based on a total available receiver capacity that is used after setting limitations on all flows. The value of $l^{(c)}$ dictates in startup stage that without affecting the RTPS mechanism, the total transmission rate of all flows works in a standard TCP manner. The purpose of LCCM is to fully utilize the available bandwidth of the network. However, due to collision of some specific flows, this initial division cannot provide full utilization of link. Therefore, RTPS increases $l^{(c)}$ gradually to achieve the maximum value. However, the increased amount is only assigned to more popular nodes, but not every node. Achievement of desired value for existing network conditions by $l^{(c)}$ and implementation of its value according to variable network environment is goal of LCCM. Intermittent re-initialization has been introduced to assure long term convergence for solving the problem of LCCM.

Basically, working of LCCM is based on two interpretations: 1) the transmission rate of the flow is reduced, when contention influences a flow, and 2) the transmission rate of the flow is enhanced, when contention disappears from the flow. The level of contention is severe when the calculated numbers of transmission rates is lesser and accordingly the inter-arrival time of data packet is increased. However, when the level of contention is reduced in the prescribed flow, it shows that the calculated transmission rate of the flow increases and the inter-arrival time of data packet is decreased.

The above methods are used to verify the condition of the network, when collision occurs and when it is terminated. Furthermore, the value of $l^{(c)}$ increases or decreases based on the contention level. Apart from the above condition, other conditions are required to work in the contention issue. The system must detect either the particular flow affected or the full link affected by contention, and LCCM should be able to respond in a case when contention leaves. In the case of a particular flow affected through contention, LCCM calculates the reduced transmission rate of this flow and then assigns its unused bandwidth to another flow. The fraction $\epsilon$ and the desired data rate $d^{(r')}$ is compared with the calculated transmission rate. Nevertheless, the process of assigning bandwidth to other flows should be based on this condition, i.e., if the calculated transmission rate for a connection is lower than a fraction, $\epsilon$, of its $d^{(r')}$. To achieve the full utilization of link, $l^{(c)}$ is gained at the rate of $d^{(r')}(1-\epsilon)$. LCCM supervises connections constantly that experiences contention. However, if the calculated transmission rate is larger than $\epsilon \times d^{(r')}$, LCCM assigns the new calculated total transmission rate to $l^{(c)}$ and notify that contention is settled. In LCCM, we assign $\epsilon = 0.7$. Additionally, to verify the status of the total link that is affected by contention, LCCM computes the condition: if $\epsilon \times d^{(r')}$ is greater than at least half of the connections' transmission rate then LCCM assumes that the full link is affected through contention. Therefore, LCCM sets by reduction in $l^{(c)}$ to the calculated total transmission rate. Furthermore, when contention is settled, LCCM will be able to calculate increment in total transmission rate. At the end of the contention, the increment in total transmission rate concludes that LCCM sets the value of $l^{(c)}$ to be equal to the new calculated total transmission rate. In the initial stage, RTPS considers some limitation on the connections and increases gradually to utilize the full link.

### B. Degree-centrality based Rate Calculation Module

A node that has maximum relationship with others in the network is called a popular or prioritized node. The scenario we discussed here is related to multiple senders connecting with a single receiver. Therefore, multiple senders reduce the availability of bandwidth on the receiver node and can contribute to congestion losses in the network. However, according to the concept of social property, the lost data packet may have higher priority in ASNETs. Therefore, to reduce the losses of a popular node's packet, we assigned bandwidth based on *degree centrality* ($d^{(c)}$) concept. Furthermore, RTPS provides least rate to every TCP connection, which also saves the data packets of less popular node rather than totally discarding them. Moreover, $d^{(c)}$ is helpful in reducing the congestion issues after delaying the acknowledgment for less popular nodes. It provides reliability to data packets of a popular node after sending an earlier acknowledgment. In

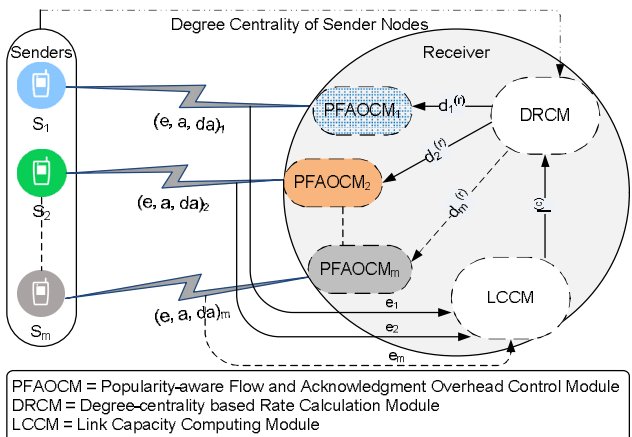

PFAOCM = Popularity-aware Flow and Acknowledgment Overhead Control Module
DRCM = Degree-centrality based Rate Calculation Module
LCCM = Link Capacity Computing Module

Fig. 2. RTPS design for overhead and reliability control in ASNETs

RTPS, we set $d^{(pop')}=1$ to distinguish the highest popular node from all connected nodes.

To calculate the popularity of a node, we used the $d^{(c)}$ concept as defined in [35]:

$$d_k^{(c')} = \frac{d_k^{(c)}}{n-1} = \frac{\sum_{i=1}^n x(k,i)}{n-1} \quad (k,i) = 1,2,3\ldots\ldots n; \quad k \neq i \qquad (2)$$

Equation (2) defines the number of direct connections between node $k$ to all other connected nodes $i$ (i.e. from 1 to $n$). To avoid the incompetence ensuing here, RTPS uses absolute value of $d_k^{(c)}$ that is equal to $d_k^{(c')}$. Therefore, in this equation, $d_k^{(c)}$ divided by $n - 1$ which is the maximum number of nodes in one social community (*SC*). The reason for designating a popular node as prioritized is its larger connectivity and more data from other nodes that needs to be transferred from one *SC* to another. Therefore, a popular node is also helpful for best utilization of resources.

To calculate the $d_k^{(r)}$, DRCM uses the degree of popularity from a sender's node and for fully utilized link capacity, it uses $l^{(c)}$ which is derived from entire flows. In RTPS, $m$ is the number of flows that shares $l^{(c)}$ bandwidth of a receiver node and $d_k^{(r)}$ represents the *k*th flow desired rate.

After calculating $d_k^{(c')}$ based rate, DRCM assigns popularity levels $d_k^{(r)}$ for PFAOCM$_k$ ($k$=1, 2… $m$). To calculate $d_k^{(r)}$, our system first assumes that the least rate is available for all sender nodes and then assigns extra bandwidth to sender nodes based on their popularity level. Nevertheless, the assignment of the least rate ($l^{(r)}$) is only possible in a situation where the value of $l^{(c)}$ is greater than all $l_k^{(r)}$. Equation (3) defines the calculation of $d_k^{(r)}$ as follows:

$$d_k^{(r)} = \frac{d_k^{(c')}}{\sum_{k=1}^m d_k^{(c')}} \times \left(l^{(c)} - \sum_{k=1}^m l_k^{(r)}\right) + l_k^{(r)} \qquad (3)$$

In this equation, $d_k^{(c')} / \sum_{k=1}^m d_k^{(c')}$ illustrates the degree centrality based assignment to $k$ flows after assigning the $l^{(r)}$ to all other connected flows. Moreover, $\left(l^{(c)} - \sum_{k=1}^m l_k^{(r)}\right)$ shows that all flows with least rates are subtracted from the total available link bandwidth. The result is provided in the form of residual bandwidth which is assigned to flows based on their popularity level.

Let us now discuss the condition under which $l^{(r)}$ cannot fulfill the requirements of the entire connected flows. In this condition the value of $l^{(c)}$ is less than or equal to all other $l_k^{(r)}$. Therefore, in this case DRCM solves this issue by initially assigning bandwidth to higher popularity level node packets. To provide the desired rate to higher popular nodes, we first arrange all popularity level in a decreasing order and then DRCM assigns the bandwidth in a decreasing order. Equation (4) computes $d_k^{(r)}$ for each connection:

$$d_k^{(r)} = \left(l^{(c)} \times \frac{d_k^{(c')}}{\sum_{k=1}^m d_k^{(c')}}\right) \qquad (4)$$

In this equation, the high popular node is served first and its rate is assigned to $d_k^{(r)}$ based on its popularity level of flow. Using the above conditions, RTPS provides maximum rates to higher popular nodes and also assigns some rates to lower popular nodes.

### C. PFAOCM Operations

As mentioned, the reliability of data packets of popular nodes and reduction in overhead are provided through assigning extra bandwidth and delaying in acknowledgment packets, respectively. PFAOCM adjusts the size of $a_k$ and $da_k$ after taking $d_k^{(r)}$ and $e_k$ as input. To achieve the desired rate based on popularity level, PFAOCM continuously sets $a_k$ and $da_k$ based on the popularity level of the node. In ad-hoc networks, the estimation of rate is too crucial due to its dynamic behavior and wireless channel interferences. Therefore, PFAOCM starts the estimation of $w_k$ after the calculation of bandwidth estimation process as shown in (1). To calculate the value of $w_k$, RTPS uses the exponential weighted moving average. Equation (5) defines $w_k$ as:

$$w_k = \sigma \times \left(w_k^{(p)}\right) + (1 - \sigma) \times e_k \qquad (5)$$

In this equation, $e_k$ defines the available rate for *k*th flow and $w_k^{(p)}$ explains the previous exponential weighted moving average. In addition, σ denotes the average parameter of exponential and RTPS sets the value of σ to 0.3. The calculation of $w_k$ and $d_k^{(r)}$ is helpful to set the value of $a$. To adjust the rate of $a$, PFAOCM waits for a minimum of one RTT after modification in $a$ or arrival of new data packets at the receiver side.

To set the value of $a$, PFAOCM considers that $a$ always uses integer number of packets and also uses Maximum Segment Size (MSS) of TCP for actual implementation. This calculation methodology is also helpful to solve congestion issues by using prioritization based partitioning of bandwidth. The calculation of $a_y$ as defined in equation (6) shows the actual estimated rate:

$$a_y = \frac{w_k \times \text{RTT}}{p_k^{(s)}} \qquad (6)$$

In this equation, $w_k$ shows the actual exponential weighted moving average rate, RTT defines the per flow estimated average round trip time and $p_k^{(s)}$ shows the size of data packets in bits that are sent from sender $k$ to the receiver. Equation (7) depicts the popularity based assignment of $a_z$ as:

$$a_z = \frac{d_k^{(r)} \times \text{RTT}}{p_k^{(s)}} \qquad (7)$$

Using this equation we calculate $a_z$ based on the desired popularity rate $d_k^{(r)}$; the values of RTT and $p_k^{(s)}$ are the same as (6). To achieve the actual advertised window size $a^{(d)}$ and actual desired rate $w^{(d)}$, PFAOCM subtracts (6) and (7).

To set the advertised window based on the popularity level, PFAOCM aims to manage $w_k$ of the flow within a desired rate $d_k^{(r)}$ of the fraction σ. The purpose of RTPS is to achieve a desired rate through $w_k \in [d_k^{(r)}(1-\sigma), d_k^{(r)}(1+\sigma)]$. The value of σ is equal to 0.3 and $a_k$ uses an integer value that must be greater than zero. In addition, in each RTT some data must be transferred. To provide the maximum bandwidth for higher popular nodes, it is necessary to compare $w_k$ with $d_k^{(r)}$. In the case where $w_k < d_k^{(r)}(1-\sigma)$ it implies that the *k*th flow is using less rates than its desired rate. PFAOCM sets the value after subtracting (6) from (7) and for consistency it multiplies $a^{(d)}$ by which is consistent factor of RTPS that is always less than 1. Therefore, in RTPS we set $\theta$ is equal to 0.7. The calculation methodology of $a^{(d)}$ is defined as follows.

$$a^{(d)} = \left\{ (d_k^{(r)} - w_k) \times \theta \times \text{RTT} \Big/ p_k^{(s)} \right\} \quad (8)$$

To avoid the above condition where $w_k < d_k^{(r)}(1-\sigma)$, PFAOCM must increase $a_k$ by one. Therefore, in (9) we compute $a_k$ using $a^{(d)}$ and $a_k^{(p)}$. Thus $a_k^{(p)}$ shows the previous advertised window of flow.

$$a_k = \left\{ max(1, a^{(d)}) + a_k^{(p)} \right\} \quad (9)$$

Furthermore, in the case where $w_k > d_k^{(r)}(1+\sigma)$, we need to reduce $a_k$ to maximize $w_k$. However, to ensure that some data packet is transferred, the value of $a_k$ must be greater than or equal to 1 during each RTT. To achieve linear decrement in $a^{(d)}$, PFAOCM uses 0.5 as in the following equation.

$$a^{(d)} = \frac{1}{2} + \left\{ (w_k - d_k^{(r)}) \times \text{RTT} \Big/ p_k^{(s)} \right\} \quad (10)$$

Using (10), PFAOCM sets the value of $a_k$ by decreasing $a^{(d)}$ from $a_k^{(p)}$. This process works continually until $d_k^{(r)} \geq a_k$. Therefore, $a_k$ is obtained as.

$$a_k = \left\{ a_k^{(p)} - a^{(d)} \right\} \quad (11)$$

To solve the overhead issue, we discuss the methodology regarding delay in acknowledgment. However, as enumerated above, to provide reliability to the data packets of popular nodes, RTPS acknowledges them earlier. Existing approaches used fixed or small $da$ for reduction in acknowledgment overhead. Nevertheless, PFAOCM uses dynamic $da$ which depends on the network situation and popularity level of the sender node. PFAOCM solves delay issues through reduction in wireless channel interference and avoids the time-out at the sender side. In PFAOCM, when any data packet arrives at the receiver, it counts the unacknowledged data packets through an $ac$ variable. After receiving data packets in an orderly manner, PFAOCM generates a single acknowledgment. The value of $ac$ is reset to 0 when $ac$ reaches $da$. To adjust $da$, the network situation provides an advantage in terms of dynamic adjustment. In a case where a receiver gets a non-sequence data packet, it is considered as a loss, paving the way for the receiver to send early acknowledgment to the sender. Moreover, earlier acknowledgment is also required if a node has higher popularity level than all other nodes, or $w_k$ is less than $d_k^{(r)}(1-\sigma)$. This early acknowledgment is helpful for the highest popular node in terms of reliable data transfer, informing the sender about losses and helping to increase $a$ to achieve a desired rate in a timely manner. PFAOCM estimates the network condition through $c$ and compares it with maximum estimation rate of flow ($me$) which shows the maximum value from all previously measured $c^{(p)}$ rates when the network was not congested.

The estimation ratio $e^{(r)}$ provides the upper and lower bound values which are based on the network condition level. The value of $e^{(r)}$ is based on comparison results between $c$ and $me*\Phi$. To detect the earlier congestion in path, we set the optimal value of $\Phi$ to 3 because a lower value for $\Phi$ generates more acknowledgment and a higher value for $\Phi$ generates fewer acknowledgments with maximum $da$. Therefore the equation defines the network state if $c$ is greater than $me \times \Phi$.

$$e^{(r)} = \frac{c - me}{c} \quad (12)$$

---

**Algorithm 1:** Pseudocode of loss or popularity based delay acknowledgment window adjustment $lpda()$.

1: Initialization.
2: Compute $w_k$ of estimated rate using Eq. (5);
3: $w_k^{(p)} \leftarrow 0$;
4: $da_k^{(p)} \leftarrow 0$;
5: Calculation of $lpda()$
6: {
7:     $\Delta_k \leftarrow w_k - w_k^{(p)}$;
8:     **if** $\Delta_k > 0$
9:        Compute $da_k$ decrement using Eq. (17);
10:     **else**
11:        $da_k \leftarrow 0$;
12:     **end if**
13:     $da_k^{(p)} \leftarrow da_k$;
14:     $ac \leftarrow 0$;
14:     Acknowledgment sent;
15:     $w_k^{(p)} \leftarrow w_k$;
16: **return** $lpda$
17: }

---

**Algorithm 2:** Pseudocode of delay window adjustment when arrival of data in an interval $ida()$.

1: Initialization.
2: $ac \leftarrow 0$, $da_k \leftarrow 0$, $da_k^{(p)} \leftarrow 0$;
3: Calculation of $ida()$
4: {
5:     **if** $ac < da_k$
6:        **if** out-of-order $|| (d^{(pop')}=1 \&\& w_k < d_k^{(r)}(1-\sigma))$
7:           Calculation of $lpda()$ from **Algorithm: 1**;
8:        **else**
9:           $ac \leftarrow ac + 1$;
10:        **end if**
11:     **else**
12:        $ac \leftarrow 0$;
13:        Acknowledgment sent;
14:        **if** $w_k \geq d_k^{(r)}(1+\sigma)$
15:           **if** $\Delta_k > 0$
16:              $da_k$ increases using Eq. (15);
17:           **else**
18:              $da_k$ decreases using Eq. (17);
19:           **end if**
20:        **else if** $w_k < d_k^{(r)}(1-\sigma)$
21:           **if** $\Delta_k > 0$
22:              $da_k$ decreases using Eq. (17);
23:           **else**
24:              $da_k \leftarrow 0$;
25:           **end if**
26:        **end if**
27:        $da_k^{(p)} \leftarrow da_k$;
28:     **end if**
29: **return** $ida$
30: }

Moreover, if $c$ is less than or equal to $me \times \Phi$ then $e^{(r)}$ is defined by (13) and selection of $\Phi$ which is equal to 3 is used to select the lower bound.

$$e^{(r)} = 1 - \Phi \quad (13)$$

After calculating $e^{(r)}$, we normalize the values between 0 and 1 through increment and decrement factor $\ddot{e}$. The factor $\ddot{e}$ approaches the value of 1 when $c$ is compared to $me$. This situation shows that there is no need to delay more acknowledgments because the path is not congested. Therefore the increment in $da$ is zero using (1-$\ddot{e}$) and $\ddot{e}$ shows closer value to 1. Moreover, if $c$ is less than or equal to $me$, then $\ddot{e}$ is equal to 0. This condition shows that the path is congested and therefore there is an increment in $da$ by 1 using (1−$\ddot{e}$). The increment and decrement factor $\ddot{e}$ is calculated as follows.

$$\ddot{e} = \frac{(\Phi - 1) + e^{(r)}}{\Phi} \quad (14)$$

We define the adjustment method of $da$ that is based on estimated condition of each flow, expiration of time, $a$, losses and popularity level. The variable $ac$ used to count the unacknowledged packets is based on the existing value of $da_k$. When data packets are in order, $ac$ increases by 1 and while it reaches $da_k$, the value of $ac$ is reset to 0 and acknowledgment packets are sent. PFAOCM first calculates the $w_k$ through (5) if the timer has not expired and $ac > da_k$. After that it compares $w_k$ to $d_k^{(r)}(1+\sigma)$ for first time arrival. If $w_k \geq d_k^{(r)}(1+\sigma)$ and the estimated rate of each flow increases, the value of $da_k$ is increased using (15). After adjustment of $da_k$, it assigns a value to the previous delayed window ($da_k^{(p)}$).

$$da_k = min(a_k, (1-\ddot{e}) + da_k^{(p)}) \quad (15)$$

PFAOCM sets $ac = 0$ and sends acknowledgment in case of losses or $d^{(pop')}=1$ or expiration of timer. The value of $d^{(pop')}$ is equal to 1 which is used to distinguish highest popular node from other nodes. The reason for calculating the expiration of the timer at the receiver side is: if the network condition is not good enough and it receives data packets in order after a very large delay. The too much delay in acknowledgment packets may cause expiration of RTO at the sender side. Therefore, to set the delay acknowledgment timer, we adopt the concept from [11].

The receiver sends early acknowledgment in case, when out of order data packets are received or the receiver acquires popular node data packets; otherwise it waits for time period $t$. The tolerance factor $f$ is used to calculate the timeout interval effectively in (16), the value of $f$ should be greater than 0. The value 2 as shown in this equation illustrates the expected arrival time of next data packets. The smoothed average arrival time ($t_k^{(s)}$) is used to estimate the expected time of arrival. Equation (16) explains the effective time-out interval $t_k$ for flow $k$.

$$t_k = t_k^{(s)} \times (2+f) \quad (16)$$

In the conditions when $ac > da_k$ and $w_k \geq d_k^{(r)}(1+\sigma)$ but the estimated average rate difference $\Delta_k$ of each flow shows decreasing order then $da_k$ is calculated using (17). Additionally, when one of the following conditions are satisfied, PFAOCM reduces $da_k$. The conditions are ($ac > da_k$ and $w_k < d_k^{(r)}(1-\sigma)$), or ($ac < da_k$ and any loss occurs or $d^{(pop')} = 1$), or (timer expires). The value of $da_k$ is reduced after calculating $\Delta_k$, which subtracts the previous average rate $w_k^{(p)}$ from the current average rate $w_k$. If $\Delta_k < 0$, the value of $da_k$ is 0. Otherwise, $da_k$ is set according to (17), if $\Delta_k > 0$. Algorithm 3 defines the adjustment rate of $da_k$ when any type of loss or expiration occurs or a popular node data packet is received. The $da_k$ defined in (17) is used to fulfill the requirement of above defined conditions. Equation (17) subtracts $(1-\ddot{e})$ from $da_k^{(p)}$ to set $da_k$. If the subtraction result is less than 0, the $da_k$ is set to 0.

$$da_k = max(0, da_k^{(p)} - (1-\ddot{e})) \quad (17)$$

The details of setting $da_k$ is illustrated in Algorithm 2 when the data packet arrives in an interval and $ac > da_k$. In the next sub-section we define the details of our RTPS algorithm to set $a$ based on the actual rate, desired rate and thus set $da$ using the network conditions through estimation rate and popularity level of nodes.

### D. RTPS Algorithm

We present a detailed algorithm of RTPS in this sub-section. Algorithm 3 uses $e_k$ of each flow to compute full utilization of

---

**Algorithm 3:** Pseudocode of RTPS for receiver windows.

1: After arrival of new data packet at receiver side
2: Calculation of $e_k$ using Eq. (1);
3: $a_k \leftarrow 1, f \leftarrow 1, da_k^{(p)} \leftarrow 0, me \leftarrow 0, c^{(p)} \leftarrow 0, k \leftarrow 1$;
4: $c \leftarrow \sum_{k=1}^{m} e_k, a_k^{(p)} \leftarrow 1$;
5: **if** $l^{(c)} \geq \sum_{k=1}^{m} l_k^{(r)}$
6:     Calculation of $d_k^{(r)}$ using Eq. (3);
7: **else**
8:     Calculation of $d_k^{(r)}$ using Eq. (4);
9: **end if**
10: // **Popularity-aware advertised window adjustment**
11: Compute $w_k$ of estimated rate using Eq. (5);
12: **if** $w_k \leftarrow [d_k^{(r)}(1-\sigma), d_k^{(r)}(1+\sigma)]$
13:     $a_k = a_k^{(p)}$;
14: **else if** $w_k < d_k^{(r)}(1-\sigma)$
15:     Compute $a^{(d)}$ using Eq. (8);
16:     Get $a_k$ after calculation of Eq. (9);
17: **else if** $w_k > d_k^{(r)}(1+\sigma)$
18:     Compute $a^{(d)}$ using Eq. (10);
19:     Get $a_k$ after calculation of Eq. (11);
20: **end if**
21: $a_k^{(p)} = a_k$;
22: //**Popularity-aware delay acknowledgment window**
23: **if** $c^{(p)} > me$
24:     $me \leftarrow c^{(p)}$;
25: **end if**
26: **if** $c > me * \Phi$
27:     Compute $e^{(r)}$ using Eq. (12);
28: **else**
29:     Compute $e^{(r)}$ using Eq. (13);
30: **end if**
31: Calculate increment/decrement factor from Eq. (14);
32: Calculate effective time out interval $t_k$ from Eq. (16);
33: **if** $t_k$ not expire
34:     Calculation of $ida()$ from **Algorithm: 2**;
35: **else**
36:     Calculation of $lpda()$ from **Algorithm: 1**;
37: **end if**
38: $c^{(p)} \leftarrow c$;

link, sets $d_k^{(r)}$ based on the popularity level (i.e., *degree centrality*) and then sets *a* based on the comparison between actual and desired rates. To set the *da*, RTPS uses network conditions from the whole network and each flow. The reliability issue for popular node is solved by setting less *da* that is also based on the network state. However, to avoid timeout at the sender side for superfluous retransmission, it is necessary to set proper *da*. We also use proper estimation to provide efficient *da*, because smaller *da* creates overhead in the network while larger *da* increases timeout at the sender side. In ad-hoc networks, wireless contention and interference affects the time of receiving data packets. RTPS increases *da* based on *a* so that the size of *da* will not increase rapidly in comparison to *a*. Therefore, we define a detailed pseudocode of RTPS in Algorithm 3, the sub-parts of which are defined in Algorithms 1 and 2. The Algorithm 2 starts when data packets arrive within time interval. In addition, the working of Algorithm 1 begins when data packets do not arrive within time interval or when it arrives from the highest popular node.

## IV. PERFORMANCE EVALUATION

In this section, we evaluate the proposed RTPS scheme and compare it with several related solutions. Let us first describe the simulation setup.

### A. Simulation Setup

In Fig. 3, the node R shows that it has maximum 1 Mbps bandwidth for sender nodes ($SA_1$, $SB_1$, $SC_1$ and $Sm_1$). Such a small available bandwidth is caused by local contentions at R. On the other hand, all other nodes have a bandwidth of 6 Mbps. For evaluation and execution, we used IEEE 802.11 as the MAC protocol. The performance of our proposed scheme is verified using OPNET simulation tool [36]. The number of intermediate node is also defined as number of hops and the range of intermediate nodes are from $I_1$ to $I_m$. The intermediate nodes utilize the concept of FIFO queuing with drop-tail method. DCF method is used to access the wireless channel in which the transmission range of our simulation scenario is 250 m. Additionally, the range of interference and the carrier sensing range of each node are set to 550 m. We varied the number of TCP connections between 1 and 25, and each node's buffer capacity is 50 packets. In order to transfer data concurrently from four hops away nodes, here we adjust each node position 200 m away from each other. The physical layer is using Direct Sequence Spread Spectrum (DSSS) and the transmission power of each node is 0.007 watts with 11 Mbps channel bandwidth. Static routing protocol with static environments is used for reducing the involvement of other losses such as wireless channel and mobility induced losses. The type of traffic is FTP. Moreover, the size of packet is 1,460 bytes and the time of simulation is 1000 seconds.

The RTPS is compared with TCP-DAAp [11] and TCP-DCA [37]. To calculate the *degree centrality* concept, we use the same sender nodes scenario as described in Pop-aware [16] which also describes how to calculate social popularity.

### B. Implementation of RTPS

In RTPS the process model of node is modified at the receiver side. The detail scenario of our implemented simulation setup is shown in Fig. 3. The process model of node works in a standard TCP manner. However, it has also additional features such as advertisement window adjustment, timestamp options and delayed acknowledgment. In order to implement the RTPS we did modification in the process model that is linked with the receiver node. The implementation of RTPS depends on two processes in which one is called parent process i.e., denoted through *tcp_manager_v3*. In addition, the second process called child process that represents as the sub-model of parent model and denoted through *tcp_conn_v3*. The finite state machine *tcp_manager_v3* communicate with the network layer and session layer. To establish a new connection the process *tcp_conn_v3* is called through *tcp_manager_v3* that is created separately for each connection. In order to work efficiently, RTPS should arrange the state information of each connection. Therefore, to handle all TCP connections, the transmission control block is used to store all separate connections parameters from *tcp_conn_v3*. Furthermore, these parameters transfer to *tcp_manager_v3* finite state machine for sharing. To implement the RTPS, we changed the *tcp_manager_v3* OPEN state that is used to call the process *tcp_conn_v3*. The adjustment of advertisement and acknowledgment window is done in *tcp_conn_v3* process. The *tcp_conn_rcv_buff_adjust()* function is used to adjust the size of receiver buffer. In order to adjust the $a_k$ value based on the source node priority level and receiver buffer size, the function *tcp_seg_receive()* is used. Therefore, we did implementation in the *tcp_seg_receive()* function to adjust the $a_k$. The *tcp_ts_info_process()* function is used to calculate the smoothed RTT value with timestamp option and the estimation period of bandwidth is simply based on the arrival of new data packets.

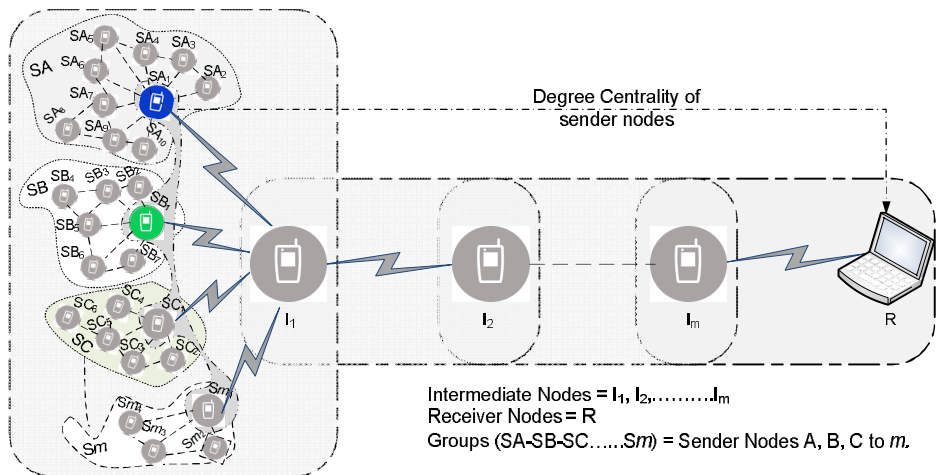

Fig. 3. Simulation setup of ASNET with multiple senders using *m* number of intermediate nodes

The implementation of RTPS is totally based on receiver side. However, to implement the full scenario we set the FIFO scheduling method at intermediate nodes. Furthermore, we adjust RED gateway with the minimum and maximum threshold sizes of packets 30 and 40 respectively. The selection of route between sender and receiver node is static that is helpful to check the performance of RTPS in terms of overhead. This is because dynamic routing creates huge amount of control packets for route discovering and selections. Basically, the aim of RTPS is to reduce the collision loss and overhead cost through reduction in acknowledgment control packets and provide end to end solution. Therefore, RTPS transfers less acknowledgment in response to each data packets.

In order to adjust the delay in acknowledgment $da_k$, we did modification in *tcp_ack_schedule()*. The adjustment of acknowledgment window is based on some conditions, in which first one is related to maximum delay time that shows how much time acknowledgment packet should delay at receiver side. To calculate the time in which the delay packet should be transferred, is made through modification in *Tcp_Fasttimo_Next_Timeout_Time_Obtain()*. RTPS handles loss after considering inter-arrival time period and when destination received data in out-of-order. To measure the out-of-order packets at receiver side, we use the *op_prg_list_size()* function. The consideration of time period at receiver side illustrates that the acknowledgment packets should transfer to source node within some specific time interval that is helpful to avoid RTO at sender side.

To understand the loss condition, let us suppose the source node sends two data packets in which one is dropped and the other one received at the destination node. After receiving single data packet, the destination node will start timer. Accordingly, when the destination node time expires for receiving data packet, then the destination node will transfer the acknowledgment packet to the sender. In the response to reception of acknowledgment packets at the sender side, the source node will transmit more data packets that create out-of-order sequence at the receiver side. This is because in previous transmission the destination node did not receive a data packet and the arrival of data packets are not in sequential manner. After receiving out-of-order data packets, the receiver node sends duplicate acknowledgments to senders and reduces the acknowledgment window. To cover-up the lost data packets the sender node retransmits them and enter into fast retransmit/recovery stage after reducing the size of congestion window to halves. Furthermore, RTPS does not transfer extra control packet for taking explicit information from an intermediate nodes or network layers. This is because, RTPS maintains end-to-end TCP semantic and does not take any explicit information related to congestion and contention status from an intermediate node.

### C. Performance Metrics

In our simulation, the performance metrics are the desired rate, throughput, latency, bandwidth division and overhead control. The desired rate of each node is computed based on the available bandwidth and the node's popularity level. To check the performance of RTPS, the throughput and transmission delay are computed as the number of hops, number of connections and packet loss rates. Moreover, to evaluate the division of data between higher, average and lower popular sender nodes, we use the bandwidth metric with time in three different topologies. Finally, to check the performance of ASNETs nodes, we also considered the monitoring and coordination overhead cost with number of connections.

### D. Performance Results and Discussions

In ASNETs, senders usually run multiple applications to communicate with each other. The criteria of sharing of scarce bandwidth resources in standard TCP are based on RTT that is not enough to fulfill the requirement of user's in ASNETs. Therefore, to share resources among users and fulfill the requirements of each sender node, we design RTPS i.e., based on receiver side. In order to deploy RTPS with standard TCP, RTPS does not enhance infrastructure of network and TCP protocol. Additionally, to maintain end-to-end semantic, RTPS also avoid taking congestion and contention notification information from an intermediate users or the other layers. Therefore, the design of RTPS is easily deployed with standard TCP protocol. To estimate the network condition in ASNETs, RTPS considered arrival rate of data packets with some specific time interval. However as compared to RTPS, TCP-DAAp deployment cost is higher because it provides both sender and receiver side solution i.e., limited to congestion window up to four and enlarge the delay window more gradually between a factor zero and one. TCP-DAAp changes the increment functionality of standard TCP, in which the sender increases congestion window up to one when it's in slow start phase and increase $1/cwnd_{i-1}$ in congestion window when it's in congestion avoidance phase. TCP-DAAp also modifies the 3 duplicate acknowledgment process and the unfairness is detected in retransmission timeout. Furthermore, in order to standard TCP, TCP-DCA breaks the end-to-end semantics of TCP. This is because TCP-DCA uses the path length information from the routing layer i.e., calculated after each hop count and adjusts delayed window based on it. To avoid delay of window greater than sender's congestion window, TCP-DCA puts congestion window information in TCP packet header.

However, as compared to standard TCP, our protocol breaks the concept of fair sharing of resources among users. This is because; in standard TCP when two flows receives same RTT it does not give any solution that limit the resources of a specific flow. Therefore, RTPS provides solution in order to avoid the reduction of highest popular node bandwidth resources. Although, RTPS still provides fairness in competing of TCP traffic from other nodes after adjusting the $a_k$ that provides limitation to the source nodes congestion window. In comparison with RTPS, the TCP-DAAp and TCP-DCA provides delay in acknowledgment window schemes and adjusts congestion window if any loss occurs that cannot provide efficient results due to limited delay and congestion window. Furthermore, these protocols cannot provide higher bandwidth resources to some specific flow. To verify the performance of RTPS with multiple aspects, we analyze in details the results in terms of degree centrality, throughput, delay bandwidth division and control overhead.

**Desired Rate:** Here we use only three sender nodes $SA_1$, $SB_1$ and $SC_1$, which have their own priority levels after calculation of *degree centrality*. According to the simulation scenario and our calculations, the result shows that $SA_1$ has the highest popularity level, $SB_1$ and $SC_1$ show decrement behavior in the popularity level. To analyze $d_k^{(r)}$ of each sender node, RTPS initially checks the available bandwidth for utilization. After that it assigns minimum bandwidth to each sender node, however, the minimum rate should not exceed the available bandwidth. Furthermore, RTPS assigns extra bandwidth to a popular node and the distribution of the remaining bandwidth is based on the popularity level. Fig. 4 illustrates $d_k^{(r)}$ values of each sender node with variable bandwidths. The minimum bandwidth requirement of each flow is set to 50 Kbps in our simulation setup. The graph shows that the availability of bandwidth varies from 200 to 1000 Kbps. First, RTPS assigns 50 Kbps to each flow and the remaining bandwidth is assigned according to *degree centrality* after inserting value in (3). The result shows that $d_k^{(r)}$ of $SA_1$ is greater than all other sender nodes and the remaining sender nodes achieve $d_k^{(r)}$ based on their own popularity level.

**Throughput:** Fig. 5(a) illustrates the overall throughput of RTPS scheme with different number of hops varying from 3 to 15. The placement of sender nodes are located on the first nodes that are denoted by *S* with different groups (A, B, C, ….., *m*). Additionally, the receiver node *R* is placed on the last node after multiple intermediate nodes ($I_1$, $I_2$,…., $I_m$). The throughput of RTPS decreases as the number of hops increases because the larger number of hops creates high contention and congestion in the network. The result shows that the performance of RTPS is better than TCP-DAAp and TCP-DCA due to lower overhead in the network. RTPS throughput is 39% greater than TCP-DAAp and 46% higher than TCP-DCA. The throughput of RTPS is efficient with maximum number of hops. This is because it uses variable (*a*) and (*da*) that depends on the popularity level of the sender node and network conditions. However, to achieve the throughput, TCP-DAAp and TCP-DCA use a maximum of four and static three delay window sizes respectively, which are not suitable with large number of hops.

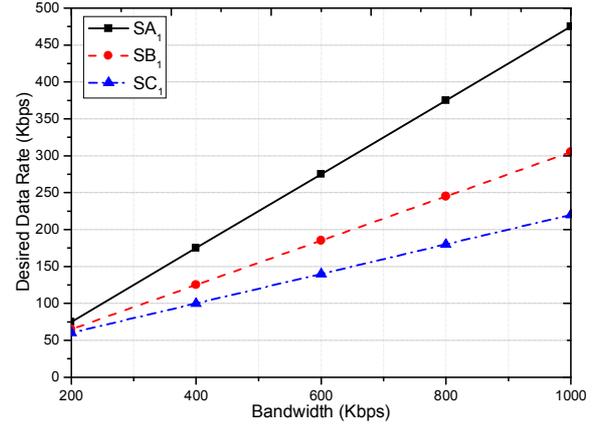

Fig. 4. Division of bandwidth among High ($SA_1$), Average ($SB_1$) and Low ($SC_1$) popular nodes.

To evaluate the performance of existing protocols, Fig. 5(b) illustrates the throughput for different number of TCP connections. The range of connections is from 1 to 25 and the numbers of hops is three. The maximum number of connections creates congestion in the network due to multiple senders sending data and receiving acknowledgment. Therefore, the throughput of RTPS shows in decreasing order due to large number of connections. However, the throughput of RTPS is 34% and 44% higher than TCP-DAAp and TCP-DCA, respectively. The reason behind higher throughput in RTPS is that it delays more in acknowledgment when the bandwidth of the network is not too high. The delay in acknowledgment is also helpful to reduce the cost of overhead in the network. To improve the throughput of RTPS we assign maximum bandwidth to that user who has a high popularity level. This is because a higher popular node has the maximum number of connections with other nodes. Therefore, it transfers maximum data from the network and achieves higher throughput. TCP-DAAp and TCP-DCA use equal bandwidth sharing with a fixed number of window sizes which cause fewer throughputs when the number of connections increases in the network.

Fig. 5(c) shows the throughput of RTPS with different loss rates. These results represent simulations of one TCP flow and 12-hop topology. The throughput of the network decreases as the packet loss rate increases. In Fig. 5(c), the comparison of RTPS with TCP-DAAp and TCP-DCA illustrates that it is 50% and 72% greater than TCP-DAAp and TCP-DCA at 10% loss rate of packets. The degradation of throughput for TCP-DAAp and TCP-DCA in high loss rate is due to high overhead and fixed window sizes within large number of hops. However, RTPS changes the *da* using condition of the network.

**Latency:** In Fig. 6(a) the latency in transmission shows that the number of hops varies from 3 to 15. The latency increases due to the increment in hops because a larger number of hops shows increment in losses. The figure shows that increment in latency pertaining to RTPS is less in comparison to TCP-DAAp and TCP-DCA. The latency RTPS is 28% lower than TCP-DAAp and 43% lower than TCP-DCA. The behavior of TCP-DAAp and TCP-DCA up to 9 hops is almost the same as RTPS due to a larger delay window for smaller hops. However, with higher number of hops the behavior of TCP-DCA is worse because it sets 3 *da* for larger hops. Moreover, TCP-DAAp behaves slowly in a worse way because it sets the delay window based on network conditions but the maximum size of window is 4. The behavior of RTPS is efficient when number of hops is larger due to the dynamic setting of *da*. In addition, it is based on estimation of the entire network conditions. RTPS also uses each flow rate level. A higher delay in acknowledgment gives shorter transmission delay.

Fig. 6(b) illustrates the transmission delay with different number of TCP connections. The range of connections is from 1 to 25 and the number of hops is three. The maximum number of connections creates delay in transmission due to multiple senders sending data and receiving acknowledgment that increase the overhead cost. Fig. 6(b) shows that the latency of RTPS is 25% and 31%, which is lower than TCP-DAAp and TCP-DCA respectively. RTPS reduces delay in transmission after assigning bandwidth based on the popularity level of a node that is directly proportional to the capacity of the sender node's desired load. Furthermore, for reduction in overhead cost the selection of *da* is dependent on the *degree centrality* social property that can be attributed to the fact that the overhead is smaller in comparison with the number of connections. Besides, RTPS also considers the bandwidth of the network to adjust the windows of a receiver node.

Fig. 6(c) shows the latency of different packet loss rates. The delay exhibits an increment behavior when packet loss rates become higher. To evaluate the delay in transmission time, we compare RTPS with other existing protocols, however the number of hops are 12. In Fig. 6(c), the comparison of RTPS with TCP-DAAp and TCP-DCA shows that it is 14% and 18% less than TCP-DAAp and TCP-DCA, respectively at 10% loss rate of packets. The minimum latency of transferring data packets in RTPS is due to network condition based delay in the acknowledgment. Multiple hops enlarge the contention level of nodes, thereby increasing the packet loss rate. To overcome this issue, it is necessary to increase/decrease *da* based on the network conditions to avoid the reliability and retransmission issues.

**Bandwidth Division:** Here we discuss the three different topologies that are useful to understand the division of bandwidth in RTPS: 1) when loss occurs at flow of highly popular node, 2) when the popular node has larger RTT as compared to a less popular node, and 3) when the overall capacity of link is reduced.

To discuss the contention related loss in RTPS, Fig. 7(a) illustrates the division of bandwidth in RTPS. Fig. 3 explains our basic simulation model and shows that $SA_1$ has the highest priority due to its maximum popularity level. Moreover, $SB_1$

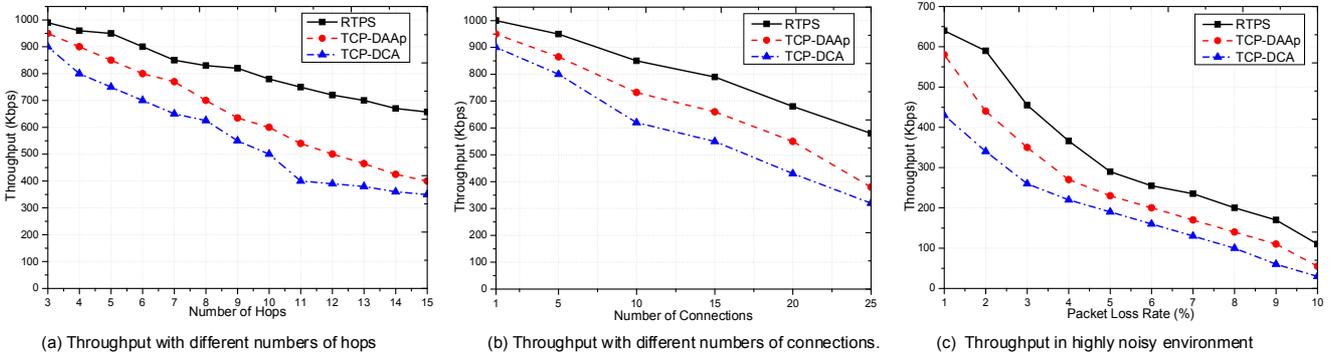

(a) Throughput with different numbers of hops　　(b) Throughput with different numbers of connections.　　(c) Throughput in highly noisy environment

Fig. 5. Performance comparison of RTPS, TCP-DAAp and TCP-DCA in terms of throughput.

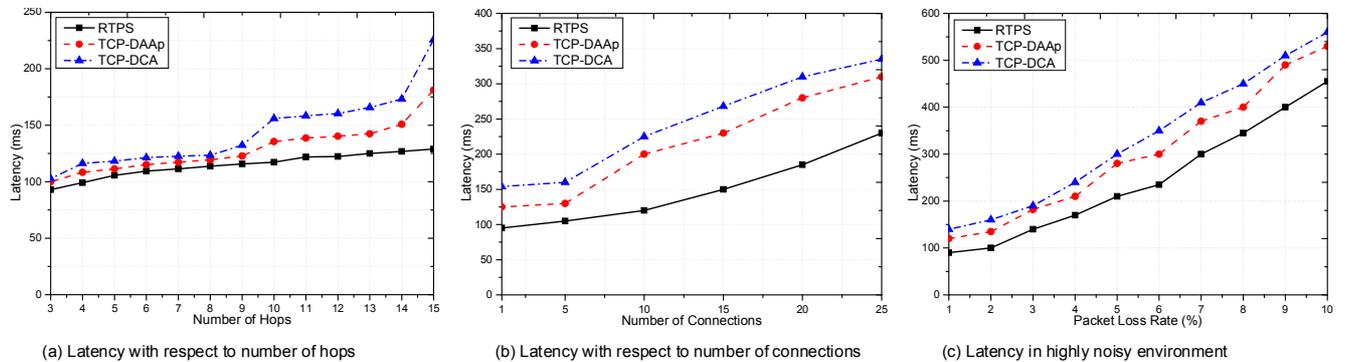

(a) Latency with respect to number of hops　　(b) Latency with respect to number of connections　　(c) Latency in highly noisy environment

Fig. 6. Latency in transmission.

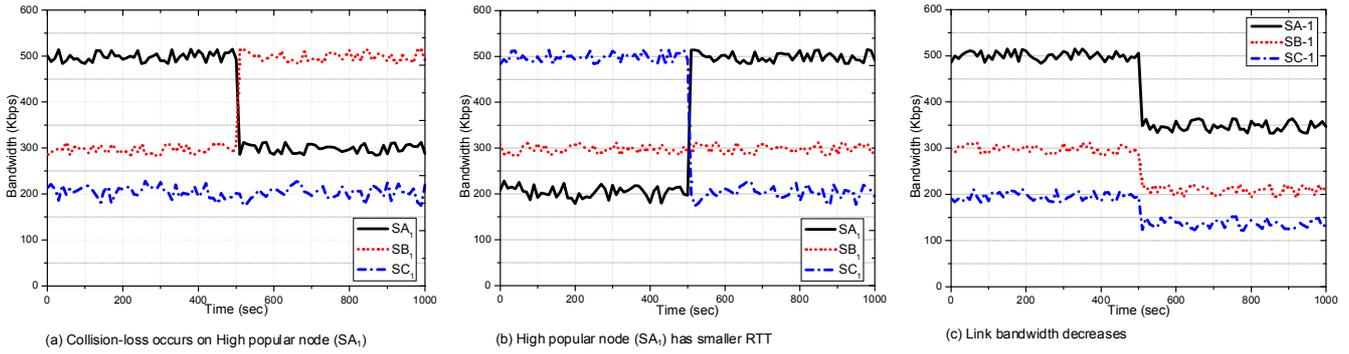

Fig. 7. Bandwidth division in RTPS after 500 seconds.

and $SC_1$ shows decreasing order in priority level. In Fig. 3 each sender and a single receiver has 1000 Kbps link capacity. To understand the contention issue, we design an extended version of existing topology in which Interference Nodes (INs) reduce the utilization of link of highest popular node. Therefore, in this topology, highest popular node waste the bandwidth and will not utilize the bandwidth more than 300 Kbps due to INs. According to this topology, RTPS reduces the wastage of resources and provides efficient utilization of bandwidth after assigning bandwidth to next popular node. To verify the performance of RTPS, Fig. 7(a) illustrates the results in which the assignment of bandwidth is based on the popularity level of nodes up to 500 seconds. After adjusting the bandwidth based on social popularity level, contention losses that occur after 500 seconds for the particular flow ($SA_1$) are considered in our simulation. Therefore, RTPS assigns $SA_1$ bandwidth to the next popular node. In Fig. 7(a), the trend shows that after 500 seconds the range of $SB_1$ bandwidth increases to 500 Kbps and $SA_1$ decreases to 300 Kbps. The result shows that RTPS utilizes efficient bandwidth when the loss occurs in a particular flow. This division of bandwidth is also helpful in increasing the throughput of RTPS to avoid wastage of bandwidth.

To understand the concept of bandwidth assignment in RTPS, we discuss the topology in which a more popular node has higher RTT while a less popular node has lower RTT. Fig. 7(b) displays the simulation result. First we assign bandwidth based on traditional TCP and the values of RTT are 330 ms, 150 ms and 50 ms for $SA_1$, $SB_1$ and $SC_1$ respectively. In the traditional TCP concept, the assignment of higher bandwidth is due to smaller RTT. Therefore, before 500 seconds $SC_1$ has a higher bandwidth. But in RTPS, the assignment criteria are the popularity level of a node. Therefore, in Fig. 7(b), 500 seconds are initially used for the traditional TCP bandwidth assignment method and after 500 seconds we run RTPS and check the assignment method. The results show that after 500 seconds, RTPS achieves bandwidth based on popularity level which is helpful in ASNET scenarios because a popular node has higher priority due to a large number of data from other nodes. Therefore, wastage of resources can occur in the network if possible bandwidth is not assigned, based on a node's requirement.

In order to verify the performance of RTPS, we discuss a topology in which the capacity of overall link is reduced. To reduce the capacity of link in this topology, we involved the UDP traffic. The reason of bandwidth link reduction in RTPS is due to non-consideration of UDP traffic for adjustment. In this topology each sender and a receiver has 1000 Kbps but the intermediate node has 6000 Kbps. After adjusting the minimum desired rate among all senders, the sender $SA_1$ has highest priority, the senders $SB_1$ and $SC_1$ has decreasing order priority. To reduce the capacity of link, 300 Kbps UDP traffic is started at node $SU_1$ at simulation time 500 sec. Fig. 7(c) illustrates that when the capacity of link is 1000 Kbps and no UDP traffic is originated then at this time each sender utilizes the full available bandwidth of link according to each sender popularity level. After 500 seconds in our simulation setup, we started 300 Kbps UDP traffic that reduces the total size of link capacity up to 700 Kbps. For fair utilization of resources based on the popularity level of each sender node, RTPS reduces the link capacity of each sender node. Therefore, Fig. 7(c) demonstrates that when UDP traffic occurs after 500 sec, each sender reduces its consumable bandwidth based on available link capacity and popularity level.

**Overhead Control:** In order to consider control overhead ratio, here we discuss the results of two conditions: 1) when source node wants to monitor that destination node is receiving data packet or not, then it receives control packets (acknowledgment packets) from destination node for verification. In this case the cost of overhead monitoring increases if destination node sends lots of acknowledgment packets to the sender's, and 2) when the source node coordinates with receiver node and it does not receives data packets with in time duration or in-order, then the source node retransmits data packets to the receivers that increases the overhead cost. These results represent simulations of 10-hop topology and bit error rate is 5% with variable TCP flows.

To discuss the overhead ratio in RTPS, Fig. 8 provides the comparison results among existing methods. In Fig. 8(a), first we discuss the results of control packet i.e., also called acknowledgment packets. In our defined scenario, the purposes of control packets are to monitor the destination node i.e., receiving data packets or not. To verify the condition of data packets at sender side, the destination node sends acknowledgment packets to the source node after receiving data packets. However, the large number of acknowledgment packets creates high overhead at the MAC layer. The estimation of control overhead ratio is calculated through number of acknowledgment received, divided by total number of data packets sent. Furthermore, when the number of connections increases, the monitoring control packets will also increase that creates high overhead in the network. According to the control overhead ratio formula, Fig. 8(a) illustrates that the number of acknowledgment packet is less in RTPS as compared to data packets. Due to the less number of acknowledgment packets in RTPS, the ratio of control overhead is also reduced in comparison with other existing methods. RTPS generates acknowledgment after estimating the rate of every single flow. The calculation of rate is helpful to estimate the condition of network. Therefore, the delay acknowledgment window in RTPS is higher and generates less number of acknowledgment packets as compared to TCP-DAAp and TCP-DCA. In order to RTPS, TCP-DAAp generates higher number of acknowledgment packets due to the maximum size of delay window which is 4. However, the overhead ratio of TCP-DCA is worst as compared to RTPS and TCP-DAAp. This is because the size of delay window in TCP-DCA is maximum three and it generates more acknowledgment packets when they are lost. The higher number of acknowledgment packets increases overhead ratio at the MAC layer and also raises the contention issue.

To evaluate the node coordination overhead cost, Fig 8(b) illustrates that when the number of connection increases the overhead ratio also rises. In ASNETs, the coordination overhead ratio depends on when fortuitous timing of TCP and MAC retransmissions results. The coordination overhead ratio can also be defined as, ratio between the number of retransmitted data packets and the total number of transmitted data packets. Fig. 8(b) explains the trends of ratio of coordination ratio among RTPS, TCP-DAAp and TCP-DCA. The reason of less coordination overhead ratio in RTPS is due to reduction in the MAC overhead by alternations in advertised window. This is because, RTPS indicates the sender formerly through advertised window that how much amount of data packets is ready to receive in the future at destination node. Furthermore, in ASNETs some senders require minimal desired rate for transmission of data packets. In order to resolve above mentioned issue RTPS adjusts minimum rate for each node connection. Therefore, the chance of loss in RTPS is less as compared to TCP-DAAp and TCP-DCA and it decreases coordination overhead cost after reducing retransmission rate of data packets.

## V. CONCLUSION

It is well known that TCP has enormous importance in wired networks for Internet communication because it supports end-to-end reliable delivery with successful congestion control and recovery. To improve the performance of TCP in ad-hoc wireless networks, many approaches have been proposed in the literature. Nevertheless, these approaches are not suitable for multi-hop ad-hoc social networks (ASNETs) where social properties are major attributes for communication.

In this paper, we have proposed a solution called RTPS for social networks in ad-hoc environments using such social property as *degree centrality* of sender nodes. A higher degree centrality shows that the node is more popular in a social network and hence it requires more reliability of data packets. Consequently, the assignment of resources and provision of earlier acknowledgment should be made in accordance. Furthermore, to control the congestion related losses, RTPS also increases delay in the acknowledgment window which is based on the network state. The selection of delayed acknowledgment window is dependent on the desired level of node popularity which is an essential part of ASNETs communication.

We compare the performance of RTPS with two state-of-the-art techniques in the literature, TCP-DAAp and TCP-DCA. The results of our simulation show that the throughput of RTPS is high using the *degree centrality* concept and avoids delay in transmission after considering dynamic delayed window size. The evaluation results also show that the assignment of resources depends on the popularity level of a node. Through the simulation results on throughput and delay,

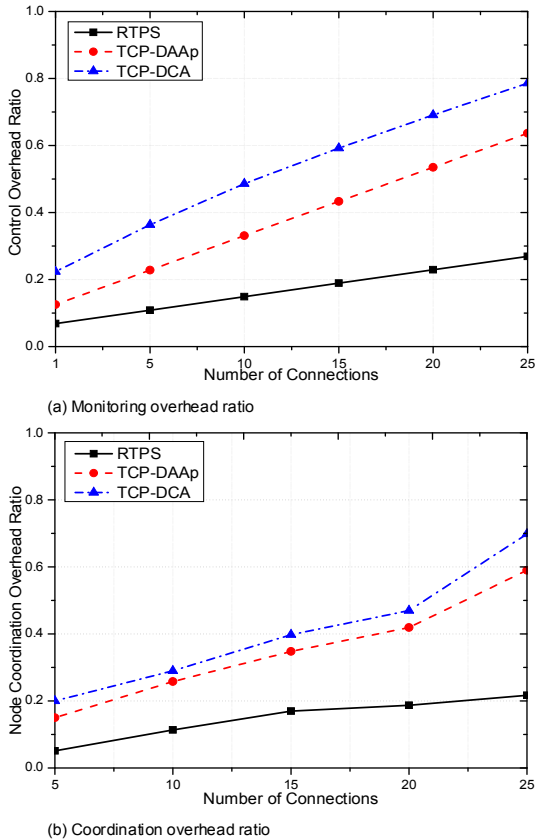

Fig. 8. Overhead ratios for node monitoring and coordination.

we show that RTPS utilizes resources efficiently compared to other schemes.

In future work, we will consider human mobility patterns and nodes capable of communicating in selfish scenarios after using dynamic routing protocol.

ACKNOWLEDGMENT

This work is partially supported by the Fundamental Research Funds for the Central Universities (DUT15YQ112, x2jq-D2154120) and the National Natural Science Foundation of China (61572106, 61572220).